\documentclass[12pt]{spieman}  
\usepackage{amsmath,amsfonts,amssymb}
\usepackage{graphicx}
\usepackage{setspace}
\usepackage{tocloft}
\usepackage{comment}

\title{On-sky speckle nulling through a single-mode fiber with the Keck Planet Imager and Characterizer}

\author[a*]{Yinzi Xin} 
\author[a]{Jerry W. Xuan} 
\author[a,b]{Dimitri Mawet} 
\author[c,d,a]{Jason Wang} 
\author[b]{Garreth Ruane} 
\author[a]{Daniel Echeverri} 
\author[a]{Nemanja Jovanovic} 
\author[e]{Clarissa Do Ó} 
\author[f]{Michael Fitzgerald} 
\author[a]{Katelyn Horstman} 
\author[c]{Chih-Chun Hsu} 
\author[a]{Joshua Liberman} 
\author[f]{Ronald A. L\'opez} 
\author[g]{Caprice L. Phillips} 
\author[h,i]{Bin B. Ren} 
\author[e]{Jean-Baptiste Ruffio} 
\author[e]{Ben Sappey} 

\affil[a]{Department of Astronomy, California Institute of Technology, 1200 E California Blvd, Pasadena, CA, 91125, USA}
\affil[b]{Jet Propulsion Laboratory, California Institute of Technology, 4800 Oak Grove Drive, Pasadena, CA, 91109, USA}
\affil[c]{Center for Interdisciplinary Exploration and Research in Astrophysics (CIERA), Northwestern University, 1800 Sherman Ave, Evanston, IL 60208, USA}
\affil[d]{Department of Physics and Astronomy, Northwestern University, 2145 Sheridan Rd, Evanston, IL 60208, USA}
\affil[e]{Center for Astrophysics and Space Science, University of California, San Diego, 9500 Gilman Dr, La Jolla, CA 92093, USA}
\affil[f]{Department of Physics \& Astronomy, 430 Portola Plaza, University of California, Los Angeles, CA 90095, USA}
\affil[g]{Department of Astronomy, The Ohio State University, 140 W 18th Ave, Columbus, OH, 43210,USA}
\affil[h]{Universit\'{e} C\^{o}te d'Azur, Observatoire de la C\^{o}te d'Azur, CNRS, Laboratoire Lagrange, F-06304 Nice, France}
\affil[i]{Universit\'{e} Grenoble Alpes, CNRS, Institut de Plan\'{e}tologie et d'Astrophysique (IPAG), F-38000 Grenoble, France}

\cftpagenumbersoff{figure}
\cftpagenumbersoff{table} 
\begin{document} 
\maketitle

\begin{center}
(Accepted July 20, 2023) \\
Submitted to Journal of Astronomical Telescopes, Instruments, and Systems
\end{center}

\begin{abstract}
The Keck Planet Imager and Characterizer (KPIC) is an instrument at the Keck II telescope that enables high-resolution spectroscopy of directly imaged exoplanets and substellar companions. KPIC uses single-mode fibers to couple the adaptive optics system to Keck's near-infrared spectrometer (NIRSPEC). However, KPIC's sensitivity at small separations is limited by the leakage of stellar light into the fiber. Speckle nulling uses a deformable mirror to destructively interfere starlight with itself, a technique typically used to reduce stellar signal on a focal-plane imaging detector. We present the first on-sky demonstration of speckle nulling through an optical fiber with KPIC, using NIRSPEC to collect exposures that measure speckle phase for quasi-real-time wavefront control while also serving as science data. We repeat iterations of measurement and correction, each using at least 5 exposures (4 with deformable mirror probes to determine phase and 1 unprobed exposure to measure the intensity), and taking about six minutes when using 59.0 second exposures, including NIRSPEC overheads. We show a decrease in the on-sky leaked starlight by a factor of 2.6 to 2.8 in the targeted spectral order, at a spatial separation of $2.0 \,\ \lambda/D$ in $K$-band. This corresponds to an estimated factor of 2.6 to 2.8 decrease in the required exposure time to reach a given SNR, relative to conventional KPIC observations. The performance of speckle nulling is limited by instability in the speckle phase: when the loop is opened, the null-depth degrades by a factor of 2 on the timescale of a single phase measurement, which would limit the suppression that can be achieved. Future work includes exploring gradient-descent methods, which may be faster and thereby able to achieve deeper nulls. In the meantime, the speckle nulling algorithm demonstrated in this work can be used to decrease stellar leakage and improve the signal-to-noise of science observations.

\end{abstract}


{\noindent \footnotesize\textbf{*}Yinzi Xin,  \linkable{yxin@caltech.edu} }

\begin{spacing}{2}   

\section{Introduction}
\label{sect:intro}
The characterization of exoplanets was identified by the Decadal Survey for Astronomy and Astrophysics 2020 as one of the top scientific priorities \cite{NRC_2020Decadal}. High-resolution spectroscopy is critical for many exoplanet measurements, including determining the planet's radial velocity, spin, atmospheric composition, as well as its surface features through Doppler imaging \cite{wang_hdc1}. It also enables the potential detection of exomoons \cite{ruffio_exomoons}.  The Keck Planet Imager and Characterizer (KPIC) is a dedicated instrument for the high-resolution spectroscopy of directly imaged companions, combining Keck II's adaptive optics system with its high-dispersion near-infrared spectrometer (NIRSPEC) \cite{delorme_kpic}. It is part of a family of instruments that leverages adaptive optics in combination with high-resolution spectroscopy, including HiRISE \cite{otten_2021}, RISTRETTO \cite{lovis_2022}, REACH \cite{kotani_2020}, and VIS-X \cite{haffert_2021}. Although the combination of high-contrast imaging and high-dispersion coronagraphy is currently unable to access rocky planets in the habitable zone, the use of this technique on the Extremely Large Telescopes has the potential to reach this regime \cite{snellen_2015}.

KPIC operates by coupling the planet's light into a single-mode fiber and routing it to NIRSPEC. Because the star is typically orders of magnitude brighter than the planet and the planet is close to the star ($<1$''), the signal on the spectrograph is often dominated by stellar light leaking into the fiber. Although cross-correlating the observed spectrum with an atmospheric model of the planet can help disentangle the star and planet signals \cite{ccf1,ccf2}, the leaked starlight nevertheless contributes a significant amount of photon noise as well as systematic errors, such as through increased telluric signals and fringing. This noise limits the sensitivity of the instrument at close separations less than 1 arcsecond, though (as a rule of thumb) KPIC becomes limited by thermal background at separations greater than $\sim 1$ arcsecond, depending on the target magnitude. 

To motivate speckle nulling, we examine the impact of stellar leakage and throughput on the integration time required to reach a given signal-to-noise ratio (SNR) on a planet's spectra. We denote the fractional planet throughput (co-axial with the SMF) as $\eta_p$ and the fractional off-axis stellar throughput (or leakage) as $\eta_s$. According to Ref.~\citenum{Ruane2018_VFN}, if the measurement is photon-noise limited (e.g. the noise scales as the square-root of photon count), the integration time $\tau$ required to reach a given SNR scales as

\begin{equation}
\tau \propto \frac{\eta_s}{\eta_p^2}.
\end{equation}

If the measurement is limited by systematic errors (e.g. the noise scales linearly with photon count), then the relative integration time to reach a given SNR scales as

\begin{equation}
\tau \propto \frac{\eta_s}{\eta_p}.
\end{equation}

In either case, it is advantageous to decrease $\eta_s$ with wavefront control, while keeping $\eta_p$ as high as possible. For wavefront control, we use a Boston Micromachines deformable mirror (DM) with 1000 actuators, which was added as part of Phase II of KPIC \cite{jovanovic_phaseii}. In this work, we explore using speckle nulling with a single-mode-fiber-fed spectrograph, adapting an algorithm originally used for suppressing starlight (speckles) on a focal-plane imaging detector \cite{bottom_spie}. Speckle nulling aims to reduce $\eta_s$ using the DM, and because the DM typically only applies a small perturbation to the wavefront, its impact on $\eta_p$ is expected to be very minor. This is unlike conventional coronagraphs, which can achieve very good starlight suppression and thus very low $\eta_s$, but may result in a sizable decrease in $\eta_p$, especially at close separations \cite{guyon_2006}.

In practice, there are other considerations that impact the relative integration time, such as different fractional overheads for each per-frame integration time, or if any frames taken for probing or calibration need to be discarded. In our case, our integration overheads while speckle nulling are only slightly higher than that of conventional KPIC observations (20\% versus 10-15\%), and, as discussed in Section \ref{sect:results}, all of the frames taken in our implementation may be used for science.

We derive the equations for speckle nulling through a fiber in Section \ref{sect:methods}, and present laboratory and on-sky demonstrations with KPIC in Section \ref{sect:results}.

\section{Methods}
\label{sect:methods}

The equations describing traditional speckle nulling at a focal-plane are given in Ref.~\citenum{bottom_spie}. We adapt those equations for speckle nulling through a fiber here.

We define the complex amplitude of the scalar electric field (``electric field" hereafter) contributed by the speckle to the focal plane as $E_{s_0}(x,y)$ and the focal plane electric field induced by the DM as $E_{\text{DM}_0}(x,y)$. The coupling of the focal plane electric field into a single-mode fiber can be described as an overlap integral between the field and the mode of the fiber centered at its physical location, denoted by (the real-valued) $\Psi(x,y)$.

We can define the (potentially time-varying) contribution of the speckle to the electric field through the fiber as

\begin{equation}
    E_s = a_s e^{i\phi_s} = \int dxdy \Psi(x,y) E_{s_0}(x,y),
\end{equation}

and the contribution of the DM through the fiber as
\begin{equation}
    E_{\text{DM}} = a_{\text{DM}} e^{i\phi_{\text{DM}}} = \int dxdy \Psi(x,y) E_{\text{DM}_0} (x,y).
\end{equation}

The electric field through the fiber can thus be described as:

\begin{equation}
    E_{\text{fib}} = E_s + E_{\text{DM}} = a_{s} e^{i\phi_{s}} + a_{\text{DM}}e^{i\phi_{\text{DM}}}.
\end{equation}

The intensity ($I_{\text{fib}}$) measured at the output of the fiber is

\begin{equation}
    I_{\text{fib}} = |E_{\text{fib}}|^2 = a_s^2 + a_{\text{DM}}^2 + 2 a_s a_{\text{DM}} \cos{(\phi_s-\phi_{\text{DM}})}.
\end{equation}

Assigning any flux through the fiber as speckle flux to be nulled, the speckle amplitude can be calculated as $a_s = \sqrt{I_{\text{fib}}}$. The speckle phase can be determined by applying DM probes. By taking measurements with $\phi_{\text{DM}} = [0,\pi/2,\pi,3\pi/2]$, we obtain the following probe measurements (assuming no noise):

\begin{align}
    I_1 &= a_s^2+a_{\text{DM}}^2 + 2 a_s a_{\text{DM}} \cos{(\phi_s)}\\
    I_2 &= a_s^2+a_{\text{DM}}^2 + 2 a_s a_{\text{DM}} \sin{(\phi_s)}\\
    I_3 &= a_s^2+a_{\text{DM}}^2 - 2 a_s a_{\text{DM}} \cos{(\phi_s)}\\
    I_4 &= a_s^2+a_{\text{DM}}^2 - 2 a_s a_{\text{DM}} \sin{(\phi_s)}.
\end{align}

The speckle phase can be estimated as

\begin{equation}
    \phi_s = \tan^{-1} \Bigl[\frac{I_2-I_4}{I_1-I_3}\Bigr].
\end{equation}

Any incoherent flux would appear in all probe measurements and thus subtract out in the phase calculation. The DM can then be used to apply $a_{\text{DM}} = a_s$ and $\phi_{\text{DM}} = \phi_s+\pi$ such that $I_{\text{fib}}=0$. Theoretically, there are many ways to achieve the desired $a_{\text{DM}}$ and $\phi_{\text{DM}}$ through the fiber. We choose to use a sinusoid on the DM, which applies a speckle with a point-source-like extent in the focal plane. We calibrate the sinusoid's spatial frequency and direction to maximize influence through the fiber, and apply  $a_{\text{DM}}$ and $\phi_{\text{DM}}$ as the sinusoid's amplitude and phase as  respectively.

This derivation assumes that the light is monochromatic. In our application, we target a narrow wavelength band (chosen to be one echelle order of the NIRSPEC spectrograph, which spans approximately $45$ nm or a $\Delta \lambda/\lambda$ of $0.02$) that corresponds to a spread in the focal plane of $\sim 0.05 \,\ \lambda/D$. Since this is much smaller than the spatial extent of a single-sinusoid speckle ($\sim 1 \,\ \lambda/D$), we expect the effect of chromaticity to be small. In Section \ref{sect:results}, we show empirically that the monochromatic assumption is indeed valid, as the wavelength band we work with is narrow compared to the null created by the sinusoid. Extending speckle nulling into broadband may be possible by calibrating multiple sinusoids that target different wavelengths, a topic left for future work.

\section{Results}
\label{sect:results}

\subsection{Laboratory Test}
A simplified diagram of KPIC is shown in Figure 1a. Light from the telescope is reflected by a deformable mirror, which is used to change the phase of the wavefront. The light then propagates through to a tip-tilt mirror, which is used to put the star's point-spread-function (PSF) on a specified pixel on the tracking camera which receives $J$- and $H$-band light from the dichroic. The $K$- and $L$-band light is sent to a focusing lens, which injects it into the SMF that then routes it to NIRSPEC.

We first tested the algorithm using Keck's internal broadband source to characterize the suppression achievable in the lab, in the absence of atmosphere. We calibrated the DM sinusoid spatial frequency and direction to maximize influence into a fiber separated by 98 mas (the predicted separation of HD 206893c at the time of testing) from the star. On KPIC, 98 mas corresponds to $2.2 \,\ \lambda/D$ in $K$-band (the science band seen by NIRSPEC, with a central wavelength of 2.2 $\mu$m). We also calibrated the amplitude of the sinusoid in DM units (from 0 to 1) to $a_{DM}$ in square-root-of-raw-contrast units.

We then perform the speckle nulling sequence, first taking the four probe frames to calculate the phase, then taking an unprobed frame to measure the raw contrast and converting it to DM units. We apply a DM correction with a gain on the amplitude (0.5 for in-lab tests, though we found that 0.25 works better on-sky) and a phase of $\phi_s+\pi$. We repeat this procedure until the null stops improving. Here, we do not apply any other perturbations to the DM and just use it to null the underlying static speckle. However, in future work, it may be possible to simulate on-sky conditions by injecting atmospheric turbulence or phase drift on the DM.

Figure 1b shows the nulling sequence, using the raw contrast (the ratio of leaked off-axis starlight to the fractional co-axial throughput) in NIRSPEC order that spans 1.99 to 2.04 $\mu$m as the metric. The ratio of initial raw contrast to the mean raw contrast of the last three iterations is 27.5. Meanwhile, because the DM-induced perturbation to the wavefront is indeed very small, the change in co-axial throughput is actually below our ability to measure, since Keck's internal light source is variable by 10-20\%. Towards the end of the nulling sequence, increasing the exposure time did not result in a deeper null, indicating that the null depth was limited by systematic effects. We did not explore the source of this limitation, since this suppression ratio was already much higher than we expected to achieve on-sky (where the performance is expected to be limited by instrumental phase drift or atmospheric turbulence). However, we speculate that this limit may be due to incoherent light in the instrument, introduced, for example, by optical elements such as dichroics, whose internal reflections can result in ghosts, or light that loses optical coherence with the main beam if the optical path difference between the two beams is greater than the coherence length. Here, the coherence length of the (combined $K$ and $L$ band) beam in question is on the order of ten microns.

\begin{figure}
\begin{center}
\begin{tabular}{c}
\includegraphics[scale=0.33]{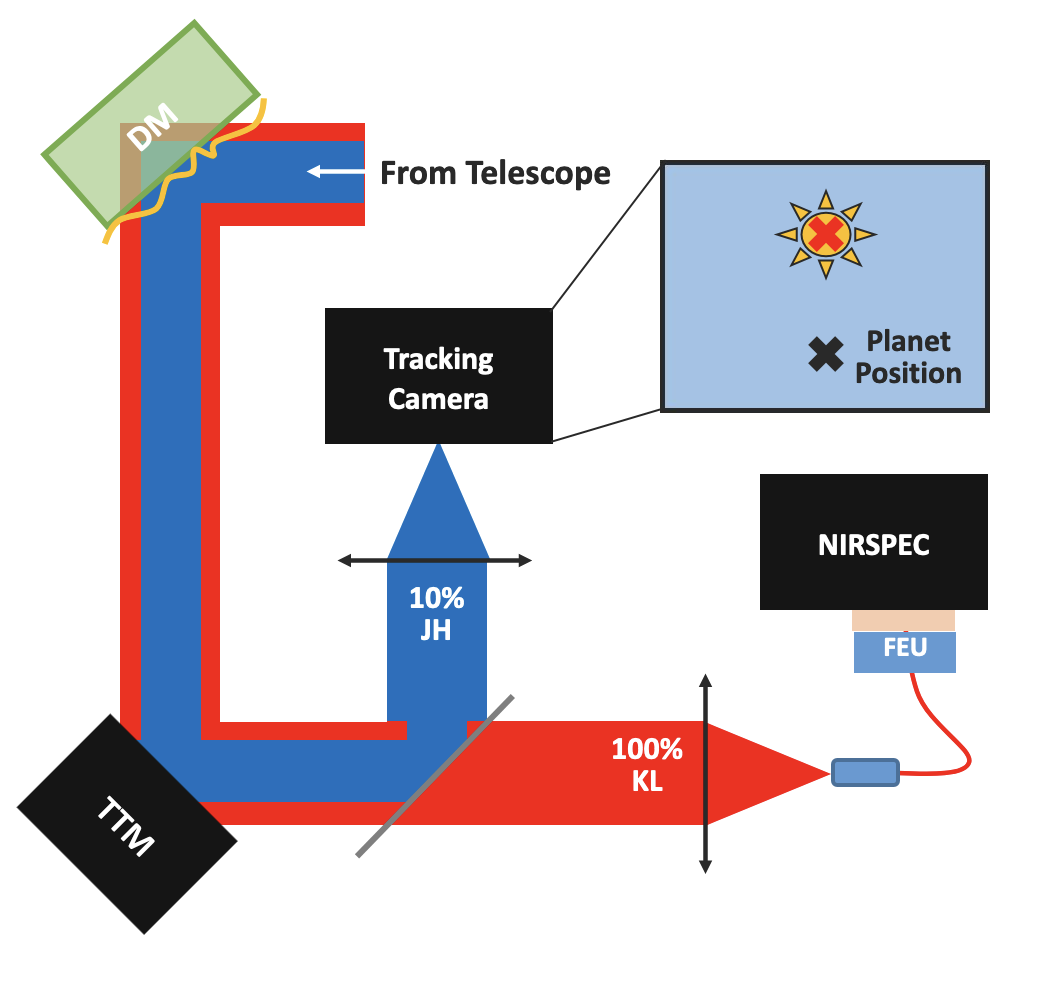}
\includegraphics[scale=0.5]{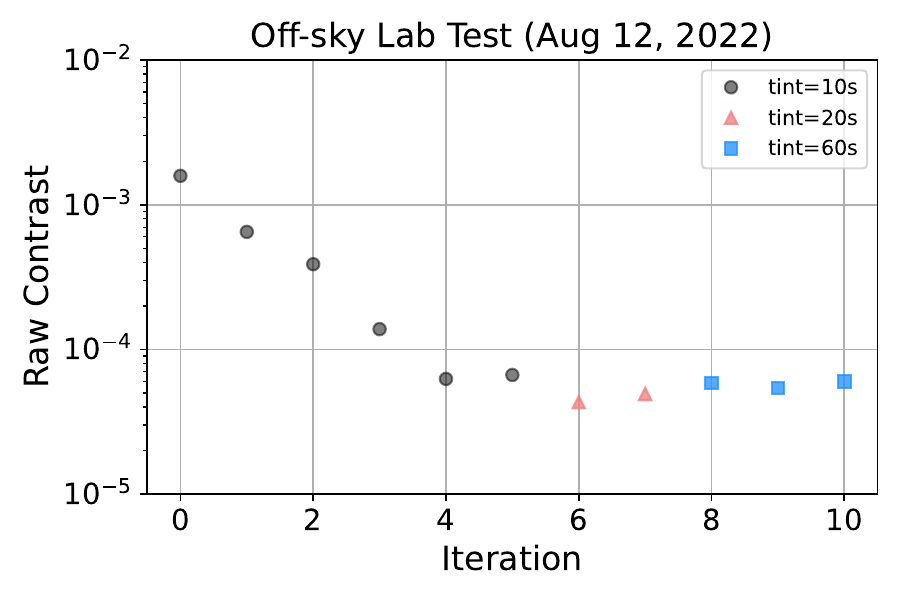}
\end{tabular}
\end{center}
\caption 
{ \label{fig:labtest} a) Diagram of KPIC. Light from the telescope is reflected off of a deformable mirror, which is used to change the shape of the wavefront. The light then propagates to a tip-tilt mirror, which is used to put the star's PSF on a specified pixel on the tracking camera that receives J and H band light from the dichroic. The K and L band light is sent to the fiber injection unit, which routes it to NIRSPEC. The inlay depicts the alignment procedure on the tracking camera: the star is positioned on the red cross such that the fiber position (indicated by the black cross) coincides with the predicted planet position. b) Raw contrast from 1.99 to 2.04 $\mu$m at each iteration during a laboratory test of speckle nulling. The ratio of initial raw contrast to the mean raw contrast of the last three iterations is 27.5. Further increasing the exposure time did not result in a deeper null, indicating that the null depth was limited by systematic effects.} 
\end{figure} 

\subsection{On-sky Engineering}

We tested speckle nulling on-sky on October 11, 2022 on HD 206893, a star with a $K$-band magnitude of 5.593 \cite{catalog}. We placed the fiber at a separation of 91 mas ($2.0\,\ \lambda/D$ in $K$-band), which was the predicted separation of the companion HD 206893c on that night \cite{witp,hinkley_2022}. The on-sky nulling sequence, targeting the raw contrast from 2.29 to 2.34 $\mu$m (where many useful CO absorption lines lie), is shown in Figure 2a. The frames were obtained using an exposure time of 59.0 seconds, with the exception of the probe measurements for iterations 6 and 7 and the raw contrast measurements for iterations 5 and 6, which were obtained with an exposure time of 119.5 seconds (to test if increasing exposure time would improve the null). There is not enough data to fully compare the behavior of the iterations using 2 minute exposures to those using 1 minute exposures, but it tentatively seems that using the longer exposure time decreases the spread in probe measurements but does not lead to a deeper null. There is a trend towards deeper raw contrasts in the beginning, but the null becomes limited around $1\times10^{-2}$.

\begin{figure}
\begin{center}
\begin{tabular}{c}
\includegraphics[scale=0.49]{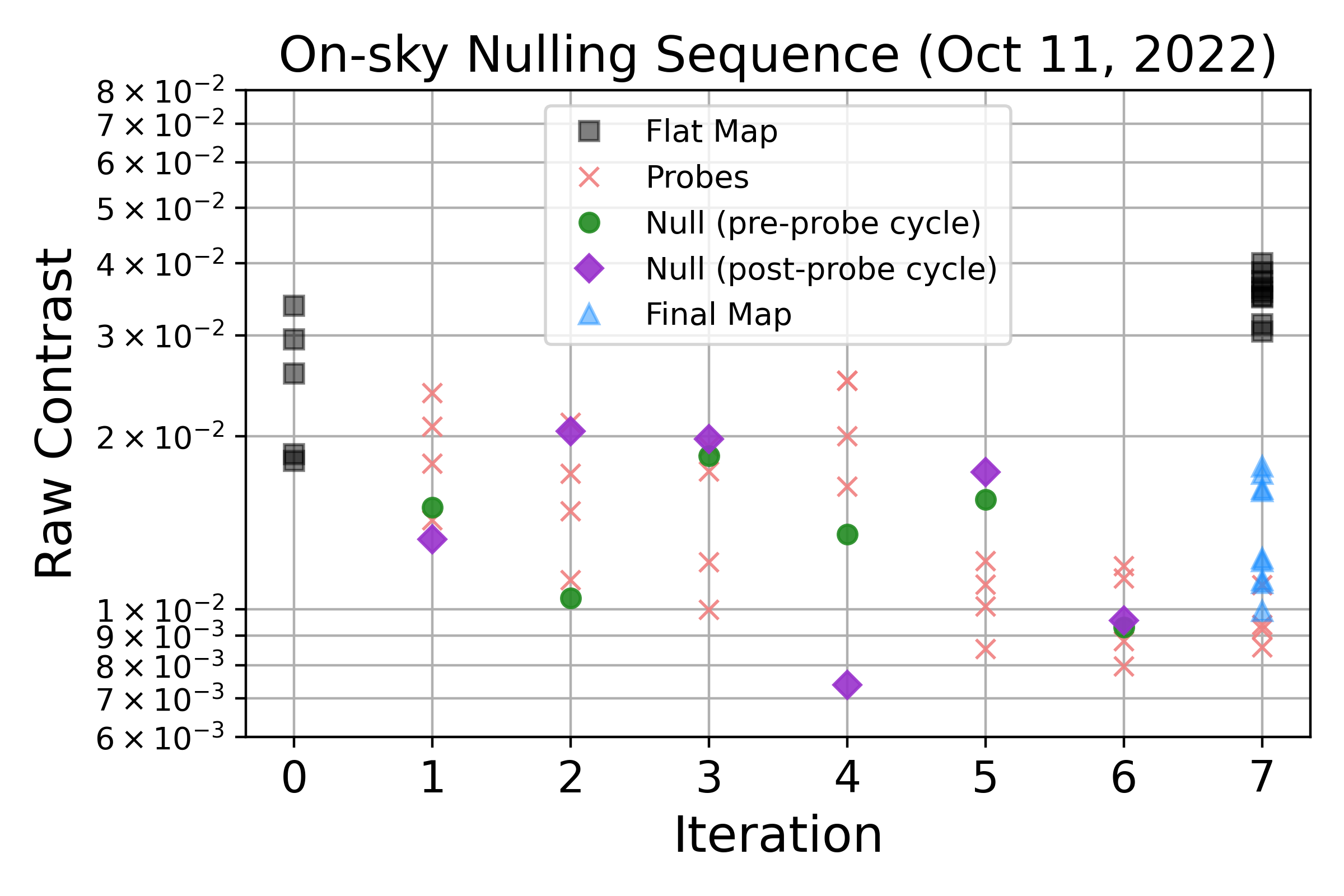}
\includegraphics[scale=0.43]{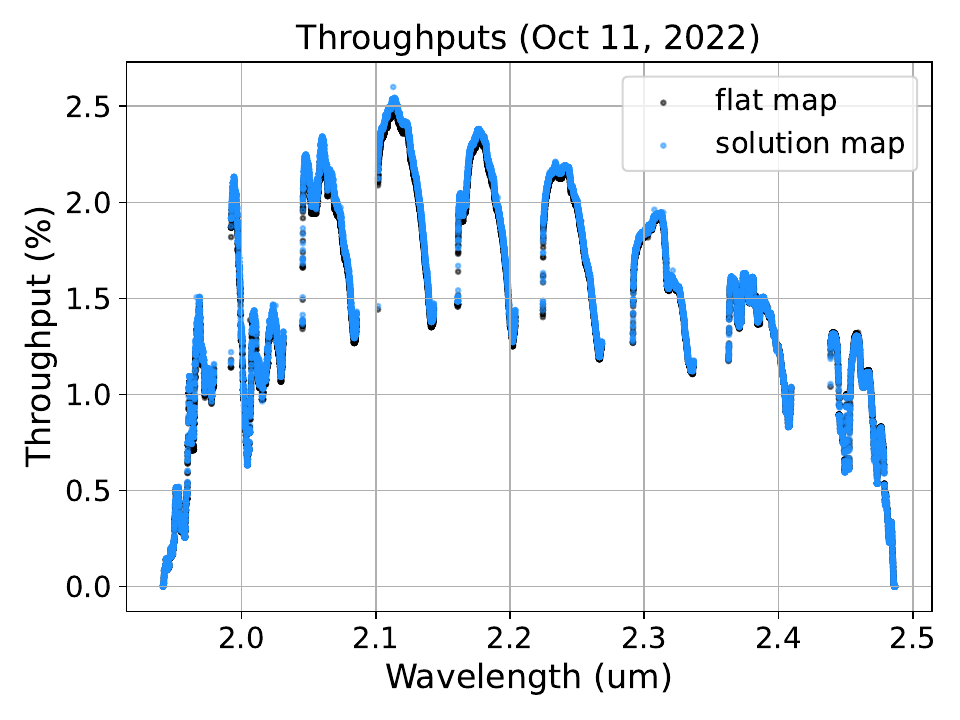}
\end{tabular}
\end{center}
\caption 
{ \label{fig:engresults}
On-sky speckle nulling results from October 11, 2022. a) Raw contrast from 2.29 to 2.34 $\mu$m at each iteration of speckle nulling. Black squares correspond to the original flat map, which is the DM map that maximizes co-axial throughput through the fiber, calibrated using the internal source before the observing night begins. Pink crosses correspond to probes used to determine the speckle's phase, as described by Equations 7-11. We take one measurement of the intermediate null immediately after applying a solution map (shown in green) and a second measurement (with the same solution map) just after the next probe cycle but before applying the next map (shown in purple). In the absence of noise or drift, we expect these measurements to be the same as the DM state is identical, so the difference between them gives us a measure of the variability. Blue triangles correspond to the final solution map at the end of seven iterations. There is significant variability in measurements with the same instrument state, but by comparing the average of the blue triangles and the average of the black squares in iteration 7, we find that speckle nulling improves the raw contrast by a factor of 2.6 relative to using the flat map. b) Comparison of co-axial throughputs with and without speckle nulling, showing that speckle nulling does not decrease the throughput $\eta_p$.} 
\end{figure} 

We observed significant variability in the measurements given the exact same instrument state. As a result, we took certain measures to characterize the level of variability, and to make sure that our comparisons and conclusions about our implementation are as valid as possible.

First, we took two raw contrast measurements of each intermediate null to measure the level of variability given the same instrument state. We take one measurement immediately after applying a solution map and a second measurement (with the same solution map) just after the next probe cycle but before applying the next map. These two separate measurements of intermediate nulls before and after the probe cycle are shown in Figure 2a in green and purple respectively.

Second, the blue triangle measurements in Figure 2a were all taken with the same DM map, and also taken as close to each other in time as possible. Although iteration 6 had a deeper null, iteration 7 is a fairer comparison to the final flat map measurements, since they were taken closer together in time, so the difference between them is less impacted by the variability.

Additionally, to check that the apparent gain from speckle nulling is not due to variability, after the last iteration, we flipped back and forth between the final DM map solution from speckle nulling and the original flat map --- i.e. we took three frames using the solution map, followed by three frames with the original flat map, then repeated this twice for a total of 9 frames with each map. Comparing the mean raw contrasts across this set of measurements, we find that speckle nulling improved the null-depth by a factor of 2.6 relative to the flat map. Additionally, the raw contrasts during the probe measurements are not noticeably higher than the measurements with the DM solutions, and are still deeper than the raw contrast without speckle nulling at all. Thus, even the probe measurements can be used as science data, and no exposures have to be excluded from analysis.

Lastly, after collecting data comparing the final solution map with the flat map, we reset the DM to our final solution map. Then, we opened the loop and took short exposures to characterize the timescale of the null degradation. The open loop raw contrast over time is shown in Figure 3. This sequence shows that the null quickly degrades --- in the time it takes to make a phase measurement using 1-minute exposures plus overheads (indicated with blue shading), the raw contrast increased by a factor of two. This degradation limits our ability to measure the drifting phase quickly enough to correct it, which is likely limiting our null-depth. This drift in phase may be due to the rotation of the pupil plane in our observation mode, as we fix the companion location to a specific point on the tracking camera and allow the pupil to rotate to maintain that position throughout the night. Unfortunately, in our case, the loop speed is limited by the photon rate, so the exposure time cannot be further decreased. Future work may involve trying gradient-descent-based approaches to nulling, or keeping a running estimate of the speckle phase. These methods may be faster at combating phase drift, since an update can be made using fewer probes. Future instruments with higher photon sensitivity may also be able to run with lower integration times and therefore at faster rates.

\begin{figure}
\begin{center}
\begin{tabular}{c}
\includegraphics[scale=0.6]{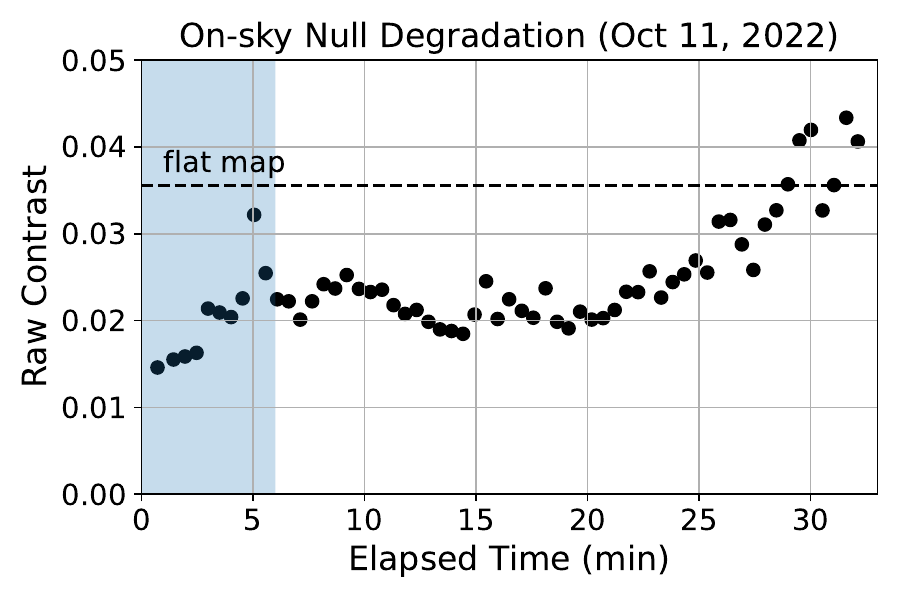}
\end{tabular}
\end{center}
\caption 
{ \label{fig:eng_phasedrift}
Raw contrast measurements over time when the loop is opened after the speckle nulling sequence. In the time it takes to make a phase measurement using 1 minute exposures plus overheads (indicated with blue shading), the raw contrast increased by a factor of two. This degradation limits our ability to measure the drifting phase quickly enough to correct it, which is likely limiting our null-depth. The dashed line indicates the raw contrast with the flat map (average of the black squares from iteration 7 from Figure 2), which is commensurate with the raw contrast that the null eventually degrades to after 30 minutes.} 
\end{figure} 

In Figure 2b, we compare the fractional co-axial planet throughput obtained with the flat map with the throughput obtained with the final speckle nulling DM solution, and show that speckle nulling does not decrease the throughput $\eta_p$. In Figure 4a, we plot the average spectrum from the last 9 frames with the flat map and the average spectrum from the last 9 frames with the final speckle nulling DM solution. We also plot the ratio of the two across wavelength to show the spectral shape of the null. The null ratios across the different orders are summarized in Table \ref{tab:nulls}. Interestingly, the null is not deepest in the targeted order from 2.29 to 2.34 $\mu$m. There is no conclusive explanation for this behavior, but one possible interpretation is that by the time these frames were taken, the underlying phase had already drifted in a way that caused the null to be deeper in an order different from the targeted one. The null is also rather broad, spanning several orders adjacent to the targeted one, and the starlight does not increase in any of the orders as a result of speckle nulling, so data across the entire observed spectrum benefits from using the technique. 

\begin{figure}
\begin{center}
\begin{tabular}{c}
\includegraphics[scale=0.48]{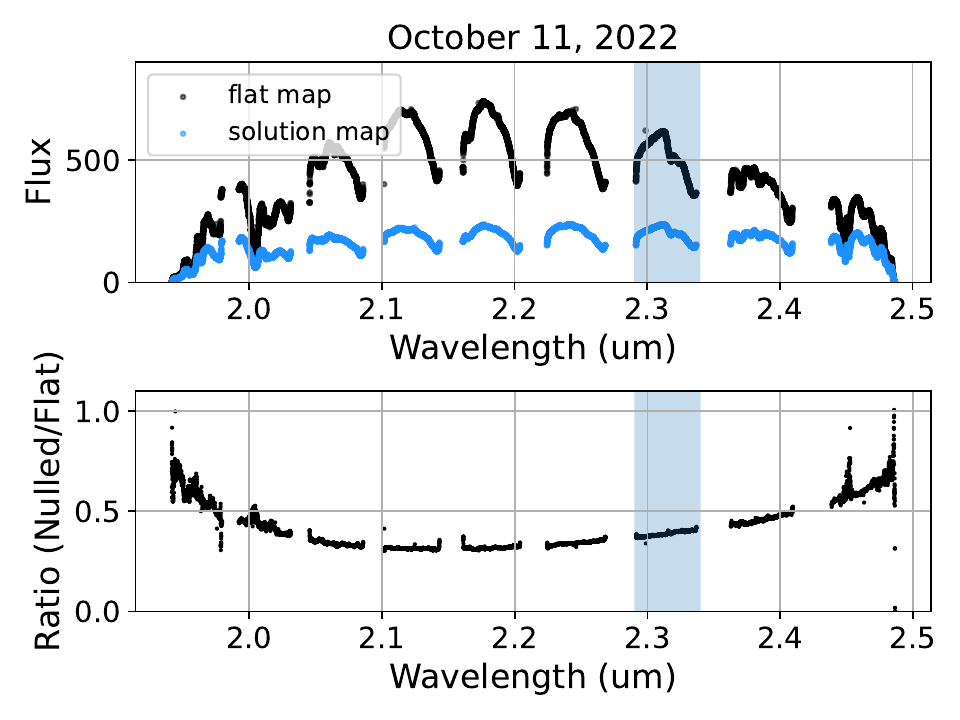}
\includegraphics[scale=0.48]{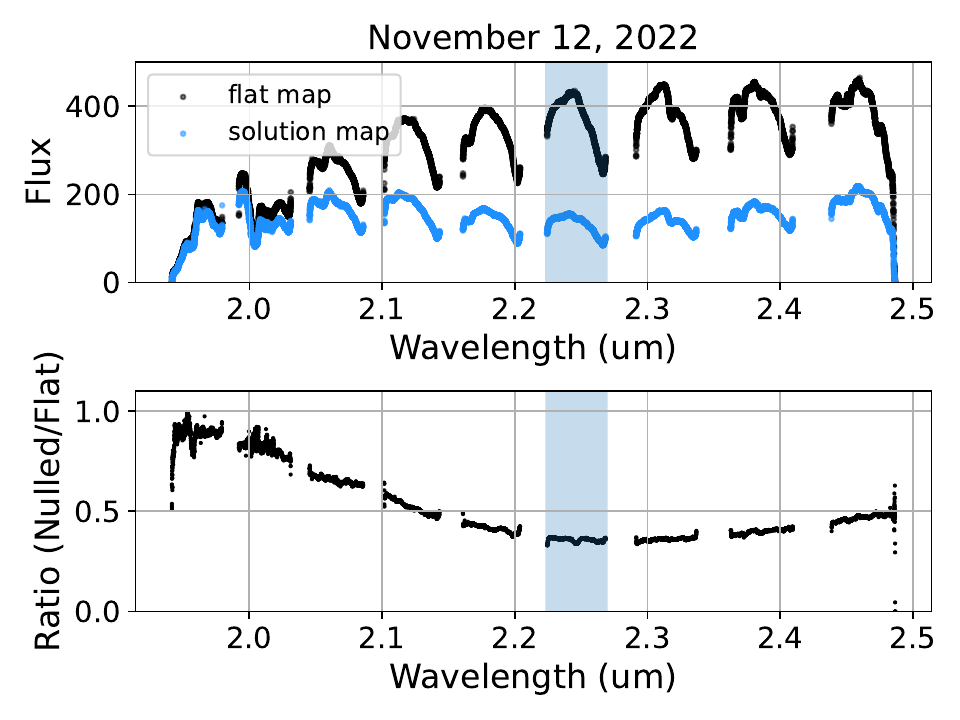}
\end{tabular}
\end{center}
\caption 
{ \label{fig:nullshape}
Comparison of speckle spectra obtained with and without speckle nulling. a) Data from October 11, 2022. On top, the average spectrum from the last 9 frames with the final speckle nulling DM solution is plotted in blue, and the average spectrum from the last 9 frames with the flat map is plotted in black. Below it, the ratio of the two shows the spectral shape of the null. The wavelength range targeted by speckle nulling (the range over which flux was summed to calculate the raw contrast) is indicated with light blue shading. In the targeted order from 2.29 to 2.34 $\mu$m, speckle nulling achieved a suppression ratio of 2.6. b) Data from November 12, 2022. On top, a frame at the end of a speckle nulling iteration is plotted in blue, and the spectrum obtained immediately afterwards with the flat map is plotted in black. Below it, the ratio of the two shows the spectral shape of the null. The wavelength range targeted by speckle nulling (the range over which flux was summed to calculate the raw contrast) is indicated with light blue shading. In the targeted order from 2.22 to 2.27 $\mu$m, speckle nulling achieved a suppression ratio of 2.8.} 
\end{figure} 

\begin{table}
\caption{Decrease in stellar coupling achieved through on-sky speckle nulling. The observed star is HD 206893, a star with a $K$-band bagnitude of 5.593. The fiber was placed 91 mas from the star at the predicted location of HD 206893c. The table lists the mean null ratio for each spectral order. Values for the order targeted by speckle nulling are in bold.}
\begin{center} \label{tab:nulls}
\begin{tabular}{ |c|c|c| } 
 \hline
 Wavelength ($\mu$m) & Null Ratio (Oct. 11) & Null Ratio (Nov. 12) \\ 
 \hline
 1.94-1.98  & 1.8 & 1.1 \\ 
 1.99-2.03  & 2.3 & 1.2 \\ 
 2.05-2.09  & 2.9 & 1.5 \\ 
 2.10-2.14  & 3.2 & 1.9 \\ 
 2.16-2.20  & 3.1 & 2.4 \\ 
 2.22-2.27  & 2.9 & \textbf{2.8} \\ 
 2.29-2.34  & \textbf{2.6} & 2.8 \\ 
 2.36-2.41  & 2.2 & 2.5 \\ 
 2.44-2.49  & 1.7 & 2.1 \\ 
 \hline
\end{tabular}
\end{center}
\end{table}

\subsection{Science Observation}

We used speckle nulling for a science observation of the same target on November 12, 2022. This time, we targeted the order from 2.22 to 2.27 $\mu$m, which has slightly higher flux on the detector, which we believed might provide slightly better signal for speckle nulling. We prioritized obtaining science-suitable exposures (including the probe measurements), so kept the loop  closed without stopping to obtain diagnostic engineering data. We only flipped back to the flat map once, immediately after a frame at the null. The comparison of the fluxes from two frames is shown in the right panel of Figure 4. The null ratios across the different orders are summarized in Table \ref{tab:nulls}. 

On this night, speckle nulling achieved a suppression ratio of 2.8 relative to using the flat map. This shows that the gain achieved with speckle nulling is repeatable over different nights.

\section{Conclusion}
\label{sect:conclusion}
We demonstrate the successful on-sky application of speckle nulling through an optical fiber, using a science spectrograph to simultaneously collect science data and measure speckle phase. We achieve a gain in stellar suppression of about 2.6 to 2.8, and show that this gain is repeatable over different nights. This suppression is achieved with minimal impact to the planet throughput (any change is well below what we can measure). Thus, using speckle nulling is expected to decrease the required integration time to reach a desired SNR by a factor or 2.6 to 2.8. The performance of speckle nulling on-sky is likely limited by the drift in phase on the timescale of a phase measurement, likely caused by instabilities in the instrument. Future work could involve exploring gradient-descent algorithms to null starlight, which may be faster and therefore better able to combat the drifting phase. In the meantime, the speckle nulling algorithm demonstrated in this work can be used to decrease stellar leakage and improve the signal-to-noise of science observations.

\appendix    

\subsection*{Disclosures}
The authors have no relevant financial interests in the manuscript and no other potential conflicts of interest to disclose.

\subsection* {Acknowledgments}
Y.X. acknowledges support from the National Science Foundation Graduate Research Fellowship under Grant No. 1122374. J.X. and D.E. acknowledge support from the Keck Visiting Scholars Program. D.E. is also supported by a NASA FINESST fellowship under award \#80NSSC19K1423. Funding for KPIC was provided by the California Institute of Technology, the Jet Propulsion Laboratory, the Heising-Simons Foundation (grants \#2015-129, \#2017-318 and \#2019-1312), the Simons Foundation (through the Caltech Center for Comparative Planetary Evolution), and NSF under grant AST-1611623. W. M. Keck Observatory access was supported by Northwestern University and the Center for Interdisciplinary Exploration and Research in Astrophysics (CIERA). The research was carried out at the Jet Propulsion Laboratory, California Institute of Technology, under a contract with the National Aeronautics and Space Administration (80NM0018D0004).

The authors wish to recognize and acknowledge the very significant cultural role and reverence that the summit of Maunakea has always had within the indigenous Hawaiian community. We are most fortunate to have the opportunity to conduct observations from this mountain.

This research made use of Astropy \cite{astropy:2022}; NumPy \cite{harris2020array}; SciPy \cite{2020SciPy-NMeth}; and Matplotlib \cite{Hunter:2007_matplotlib}.

\subsection* {Code, Data, and Materials Availability}
An abridged version of the code and data used in this work can be found at \url{https://github.com/yinzi-xin/kpic_sn_jatis}, or with the DOI: 10.5281/zenodo.8147323.


\bibliography{report}   
\bibliographystyle{spiejour}   


\vspace{2ex}\noindent\textbf{Yinzi Xin} is a graduate student at Caltech, where she works in the Exoplanet Technology Lab under the guidance of her advisor, Dimitri Mawet. Her research interests lie in the field of high contrast imaging instrumentation and data analysis for exoplanets. She is interested in wavefront sensing and control, coronagraphy, and the development of new instrument concepts. 

\vspace{1ex}
\noindent Biographies and photographs of the other authors are not available.

\end{spacing}
\end{document}